\documentclass[lettersize,journal]{IEEEtran}
\usepackage{amsmath,amsfonts}
\usepackage{algorithmic}
\usepackage{siunitx}
\usepackage{array}
\usepackage[caption=false,font=normalsize,labelfont=sf,textfont=sf]{subfig}
\usepackage{textcomp}
\usepackage{stfloats}
\usepackage{url}
\usepackage{verbatim}
\usepackage{graphicx}
\def\BibTeX{{\rm B\kern-.05em{\sc i\kern-.025em b}\kern-.08em
    T\kern-.1667em\lower.7ex\hbox{E}\kern-.125emX}}
\usepackage{balance}
\begin{document}
\title{Design of a Six-band, 2.4-Octave (80--420 GHz) Hierarchically Summed Phased-Array Slot-Dipole Antenna Array for NEW-MUSIC}
\author{Xiaolan Huang, Shibo Shu,  Miao Li, Sunil R. Golwala, Feng Liu
\thanks{{This work was funded by the project of Professor Liu Feng from Shanghai Normal University through grants from the National Natural Science Foundation of China (Grants No. 12141303, 12073018, and 12373099) and the Shanghai Outstanding Academic Leaders Program (Grant No. 22XD1422100).}}
\thanks{Xiaolan Huang is with the Shanghai Normal University, Shanghai, China (e-mail: huangxl@ihep.ac.cn).}
\thanks{Shibo Shu is with the Institute of High Energy Physics, Chinese Academy of Sciences, Beijing, China (e-mail: shusb@ihep.ac.cn).}
\thanks{Sunil R. Golwala is with the California Institute of Technology, Pasadena, California, USA (e-mail: golwala@caltech.edu).}

\thanks{Feng liu is with the Shanghai Normal University, Shanghai, China (e-mail: fliu@shnu.edu.cn).}}

\markboth{Transactions on Applied Superconductivity, ~Vol.~ , No.~ ,  ~2025}%
{How to Use the IEEEtran \LaTeX \ Templates}

\maketitle

\begin{abstract}
The Next-generation Extended Wavelength Multi-band Sub/millimeter Inductance Camera (NEW-MUSIC), located on the Leighton Chajnantor Telescope (LCT), will be the first six-band trans-millimeter wave polarimeter. This paper proposes a broadband, hierarchical phased-array antenna with integrated band-defining filters necessary to realize NEW-MUSIC. It covers a spectral bandwidth of 2.4 octaves from 80~GHz to 420~GHz, a frequency range ideal for studying trans-millimeter emission from a range of time-domain sources, using the Sunyaev-Zeldovich effects to study hot plasmas in galaxy clusters and galaxies, and to observe dusty sources, from star-forming regions in our galaxy to high-redshift dusty, star-forming galaxies. To achieve these goals, three groups of superconducting lumped-element on-chip low-pass/band-pass filter-banks were designed to hierarchically sum the superconducting, broadband, non-resonant, slot-dipole antenna arrays and band-pass filter the trans-mm light before outputting it on microstripline to detectors (KIDs in the case of NEW-MUSIC).
\end{abstract}


\section{Introduction}
\IEEEPARstart {W}{ith} the advancement of sub/millimeter wave instrumentation, the analysis of the origin of some of the most energetic phenomena in the universe can be conducted based on multi-band spectral energy distribution data across a broad spectral range. 
Among these, the complex spectral features of the Sunyaev-Zeldovich effect and foreground source contamination require multi-band observations in the trans-millimeter wave range. 
Also, simultaneous multi-band observations will benefit the study of trans-millimeter-wave time-domain sources.
The Next-generation Extended Wavelength Multi-band Sub/millimeter Inductance Camera (NEW-MUSIC) on the Leighton Chajnantor Telescope (LCT) on Cerro Toco in Chile will address these scientific needs using a focal plane covering six spectral bands over 2.4 octaves, 80 -- 420 GHz.  This extremely broad spectral response will be provided by hierchically summed phased arrays of broadband, superconducting, slot-dipole antennas. The frequency selection needed for the summing, and to define detector spectral response bands, will be done using superconducting, on-chip, lumped-element low-bass/band-pass filter banks. Such multi-band detector arrays would also enable the maximal use of focal plane area for upcoming 30 m to 50 m class trans-mm and sub-mm telescopes, for which the expense per unit focal plane area is substantially higher than current telescopes.

Hierarchical antennas, first proposed in \cite{2002AIP, 2003ISOP} and demonstrated in \cite{2018apl, JLTP2024}, address this requirement through coherent summation in a frequency-selective manner using low-pass/band-pass filter banks. The frequency scaling capability relative to pixel dimensions was first identified in \cite{2006ISOP, 2014ISOP}.
Building upon this foundation, \cite{2014ISOP} first introduced the concept for a three-scale hierarchical phased-array slot dipole antenna, which covering 75–415 GHz, coupled to $TiN_x$ KIDs, and fabricated a two-scale version with six bands over this frequency range. 
$TiN_x$ deviates from Mattis–Bardeen theory \cite{Mattis-Bardeen Theory} necessitated material substitution, leading Shu et al. \cite{jltp2022} to revised the design to use Al KIDs. Due to issues in the design of initial antenna and band-pass filter (BPF) that led the scope was reduced to four bands for an initial demonstration (without reducing the antenna’s intrinsic bandwidth) and the latest development of this project is presented in\cite{2024spie}\cite{JLTP2024}.

We present the design and simulated performance of a three-scale, six-band hierarchical antenna with integrated filter-banks for NEW-MUSIC.

\section{Phased Array antenna design}

\noindent Our hierarchical antenna array design was first introduced in \cite{2024spie}.  Here we provide more technical detail on the design considerations. The non-resonant fundamental slot antenna has a slot size of \SI{1664}{\um}$\times$ \SI{18}{\um}. Each slot has 16 feeds equally spaced along the slot by \SI{104}{\um}.The ground plane is niobium, and the feeds consist of \SI{1}{\um} wide microstripline crossing the slots terminated in shunt capacitors. The two metal layers are Niobium (\SI{190}{\nm} and \SI{160}{\nm}) and the microstripline dielectric is \SI{1070}{\nm} of low-loss, hydrogenated amorphous silicon ($\alpha$-Si:H) dielectric\cite{2024TLS noise of Si, 2024Phys. Rev. Materials}. 16 such slots are arrayed with a slot-to-slot distance of \SI{104}{\um}, and all 256 feeds are connected to a hierarchical summing tree, to form the 16 $\times$ 16 level 0 (fundamental) hierarchical array element(Fig.\ref{antenna array and impedance.pdf}(a)).

\begin{figure}
\centering
\includegraphics[width=3.5in]{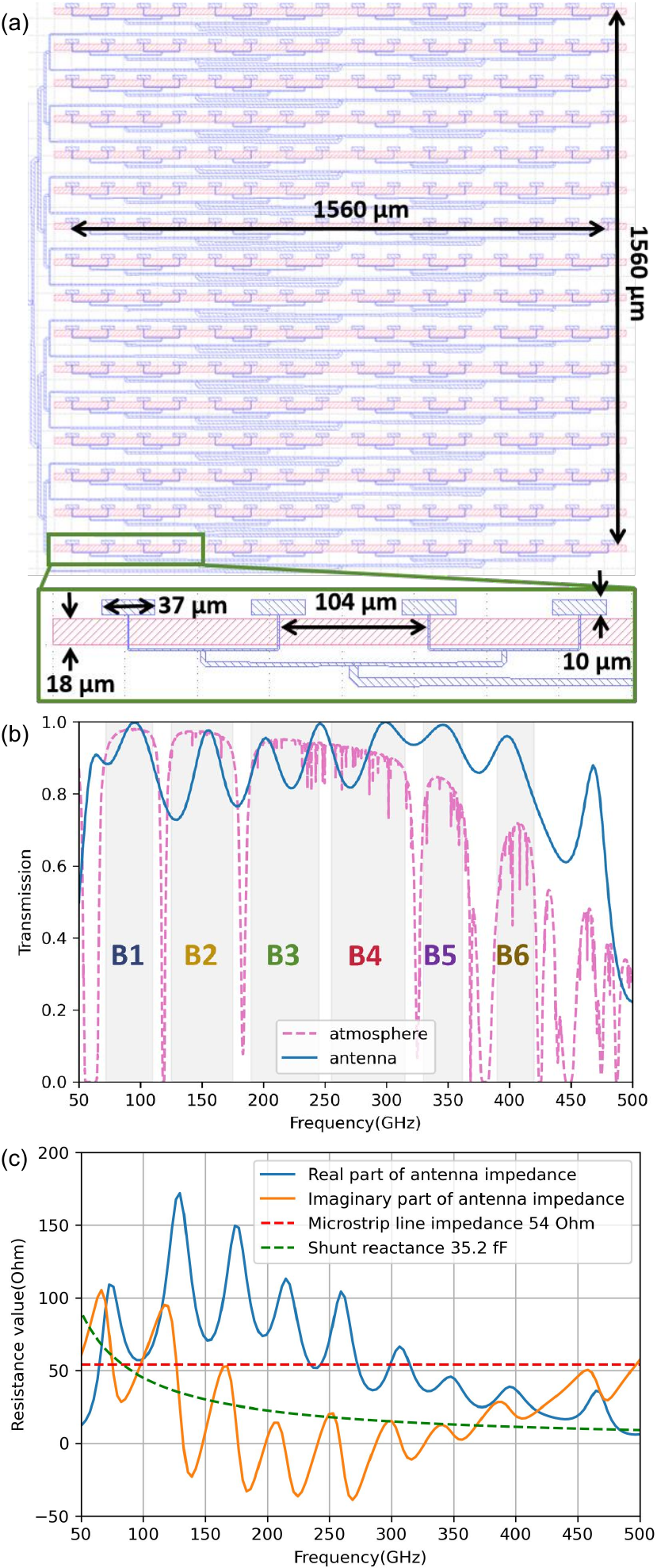}
\caption{(a) A level 0 (fundamental) element of our hierarchical antenna, composed of 16 slots each with 16 feeds. (b) The radiation efficiency of the our hierarchical antenna overlaid with the typical atmospheric transmission on the Chajnantor Plateau for 1 mm precipitable water vapor (PWV) from the ATM model*. The efficiency is for an infinite array of slot antennas without bandpass filters but accounts for the backshort and anti-reflection treatment. (c) The real and imaginary parts of the antenna's impedance ($Z_{out}$) as a function of frequency.}

{\scriptsize* \url{https://www.apex telescope.org/sites/chajnantor/atmosphere/transpwv/index_ns.php}}
\label{antenna_array_and_impedance.pdf}
\end{figure}

The antenna feed impedance, shown in Fig.\ref{antenna array and impedance.pdf}(c), is calculated numerically assuming an infinitely long slot. The microstrip feedline width is \SI{1}{\um} is chosen to provide a characteristic impedance of \SI{54}{\ohm} for impedance matching to the slot. A \SI{37}{\um}$\times$\SI{10}{\um} rectangular capacitor is added at the end of the feedline providing \SI{0.0352}{\pF} (\SI{45.25}{\ohm} at \SI{100}{\GHz}) to compensate the imaginary part of the antenna feed impedance at low frequency. The radiation enters through the substrate, which is high-resistivity silicon.  Its high dielectric constant ($\epsilon_r = 11.7$) necessitates anti-reflection treatment.  We assume a three-layer AR stack for these calculations, and we place a backshort 150 um away from the metal side of the wafer to improve efficiency~\cite{doi:10.1117/12.926055}.  Multi-layer antireflection structures using etched silicon have been demonstrated~\cite{doi:10.1364/AO.57.005196, doi:10.1109/TTHZ.2025.3555418} and will be used in practice. In our calculation, a three-layer stack is used and a backshort is placed at \SI{150}{\um} away from the top of the chip. In practical, we will utilize a multi-level antireflection layer for observations.

The feedline network has to be carefully designed within the limited space between slots. Simple T-junctions with impedance matching are used for summing signal coherently. In total, the summing tree for the $16\times 16$ level 0 element incorporates 8 levels of such T-junctions. The reflections $S_{11}$ of all T-junctions are below \SI{-20}{\dB} over the planned frequency range, and six of them are optimized below \SI{-35}{\dB}. To keep the transmission lines narrow enough for the summing tree for a single slot to fit between the slots, the lines are linearly tapered between junctions, also with $S_{11} < -20 dB$. Sections of the summing tree with long runs of microstriplines in parallel are a concern due to potential mutual coupling~\cite{doi:10.1088/0004-637X/812/2/176}. A \SI{400}{\um}-long section of two parallel transmission lines, which is the longest parallel section in our network, is optimized to have the edge-to-edge distance of \SI{8}{\um}. The crosstalk is below -20 dB when the input ports are at opposite ends.  (It is $< -38$ dB when on the input ports are at the same end.)

\section{Multi-band hierarchical antenna design}
\noindent While the antenna is broadband, we optimize the filter-bank design for six specific spectral bands with the following names and center frequencies: B1 (\SI{90}{\GHz}), B2 (\SI{150}{\GHz}), B3 (\SI{230}{\GHz}), B4 (\SI{290}{\GHz}), B5 (\SI{350}{\GHz}), and B6 (\SI{400}{\GHz}). The level 0 (fundamental) $16\times 16$ array is the pixel for B5 and B6. Four level 0 pixels together form a level 1 pixel for B3 and B4. Four level 1 pixels form a level 2 pixel for B1 and B2. Filter-banks are do the summing in this frequency-selective fashion (and define bands for detectors).

\begin{figure}
\centering
\includegraphics[width=3.5in]{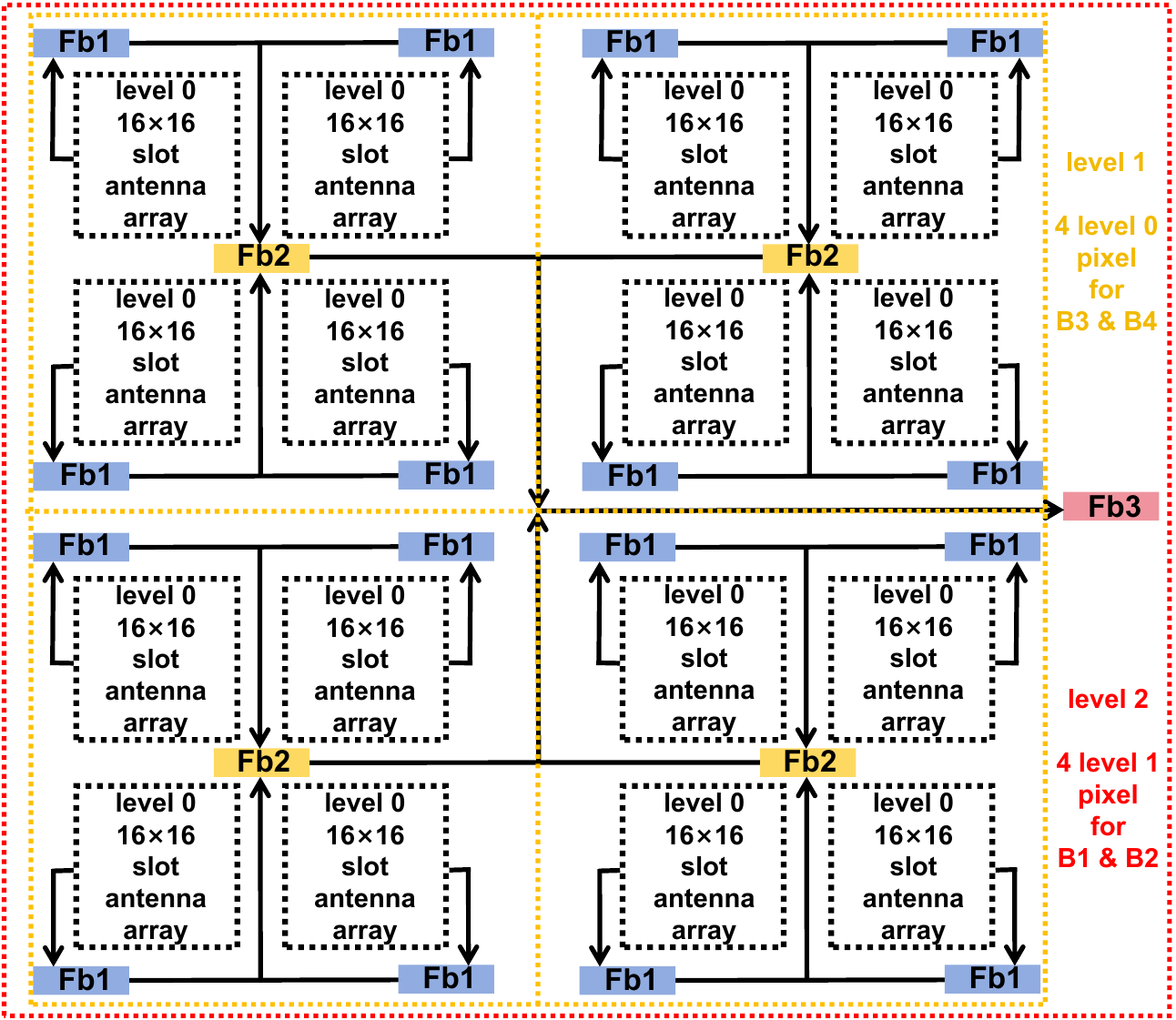}
\caption{The schematic of the three-scale  hierarchical antenna pixel design with 6 bands.  Each square unit represents a 16 × 16 fundamental element (level-0)l and each "FbX" represents a filter-bank. Of the four elements of a level 1 group, the left ones match the schematic in Fig. 1(a) while the right ones have been mirrored through a vertical line so the summing tree exits to the right, not the left. (There is no mirroring through a horizontal line because that would cause a $180^\circ$ phase shift of the feed excitation.) Fb1 consists of LPF1234 (316 GHz), BPF5 (335-360 GHz) and BPF6 (390-411 GHz). Fb2 consists of LPF12 (175 GHz), BPF3 (201-246 GHz) and BPF4 (270-310 GHz). Fb3 consists of BPF1 (77.5-106 GHz) and BPF2 (133.5-172.5 GHz). The transmission lines of the filter-banks extend to the edge of the hierarchical antenna. }
\label{hierarchical_antenna_pixel.pdf}
\end{figure}

We provide here and in Fig.\ref{hierarchical antenna pixel.pdf} an overview of the filter-bank network, with detailed designs to be presented in the following section. At the output of the level-0 pixel, there is a filter-bank ("Fb1") with band-pass filters to define B5 and B6 and a low-pass filter L1234 with a cut-off frequency of \SI{315}{\GHz} to route light to be summed for B1, B2, B3, and B4. The L1234 from two level 0 pixels are first binarily summed, and two such summed signals then enter a 5-port filter-bank ("Fb2"), which incorporates two band-pass filters to define B3 and B4 and a low-pass filter L12 to route the light for B1 and B2. The L12 outputs of four level 1 pixels are binarily summed twice using two levels of T junction and then reach a filter-bank ("Fb3") with band-pass filters to define B1 and B2. All the band-pass filters output go to and terminate in KIDs. This filter-bank scheme brings all the signals to detectors without any other cross-overs.  The long slots are broken, but we have previously shown that such breaks introduce no gross beam features\cite{JLTP2024}.

\section{Filterbank designs}
As mentioned above, three filterbanks are required to define six passbands, where low-pass filters and bandpass filters are included.
Our filterbank design proceeds in three steps.  We begin with a lumped-element circuit design and choose component values to provide the desired cutoff frequency (or frequencies) and pass-band ripple. We then define a physical layout and simulate it in Sonnet, iterating the design to obtain acceptable performance. 
Compared with distributed-type filters, adding multiple lumped-element filters together as a filterbank does not require specific distances.
We then combine the layouts for the filters in each bank, with as short lengths of transmission line between filters as possible: while each filter has a high out-of-band impedance, these short lengths act as impedance transformers, causing them to load one another.  We re-simulate in Sonnet and again iterate the design to obtain acceptable performance.
The simulation setup in Sonnet is \SI{150}{\nm} thick Nb ground layer with surface inductance of \SI{0.1}{\pH}, \SI{1070}{\nm} thick $\alpha$-Si dielectric with dielectric constant of $11.5$, and \SI{450}{\nm}-thick Nb wiring layer with surface inductance of \SI{0.05}{\pH}.

\subsection {Low-pass Filter Designs}

There are two types of low-pass filter prototypes: one begins with a shunt capacitor, and the other one begins with a series inductor. When filters are connected together as a filterbank, the shunt capacitor affects the performance of the other filters dramatically. Therefore, our LPFs begin with series inductors. 

The narrow frequency separation between frequency bands requires LPFs having fast roll-offs. We choose a 5th-order \SI{0.5}{\dB} pass-band ripple Chebyshev filter with $N = 9$ elements.
The cutoff frequencies are \SI{175}{\GHz} and \SI{316}{\GHz} for LPF12 and LPF1234, respectively. The transmission line impedances are \SI{19}{\ohm} and \SI{37}{\ohm} for LPF12 and LPF1234, respectively. This impedance is matched to the microstripline exiting the antenna for LPF1234, but it must be reduced to half that value for LP12 because the BPF3-BPF4-LP12 filterbank is also s a binary summing junction. All component values are calculated via routine LPF design method using the desired cutoff frequencies and pass-band ripple~\cite{Microwave Engineering}. The transmission of the lumped-element circuit design is calculated numerically to verify the design. Each inductor and capacitor is first individually simulated in Sonnet to ensure their values match the circuit design at the cutoff frequency. The individual components are then combined together following the LPF circuit design and simulated in Sonnet.  We find the cutoff frequency tends to move down, suggesting the influence of parasitic contributions.  We adjust the shunt capacitor sizes to match the desired cutoff frequencies. The simulated performance of LPF12 and LPF1234 is shown in Fig.~\ref{LPF1234 & LPF12.pdf}(d).

\begin{figure}
\centering
\label{tab:lpf_design}
\includegraphics[width=3.5in]{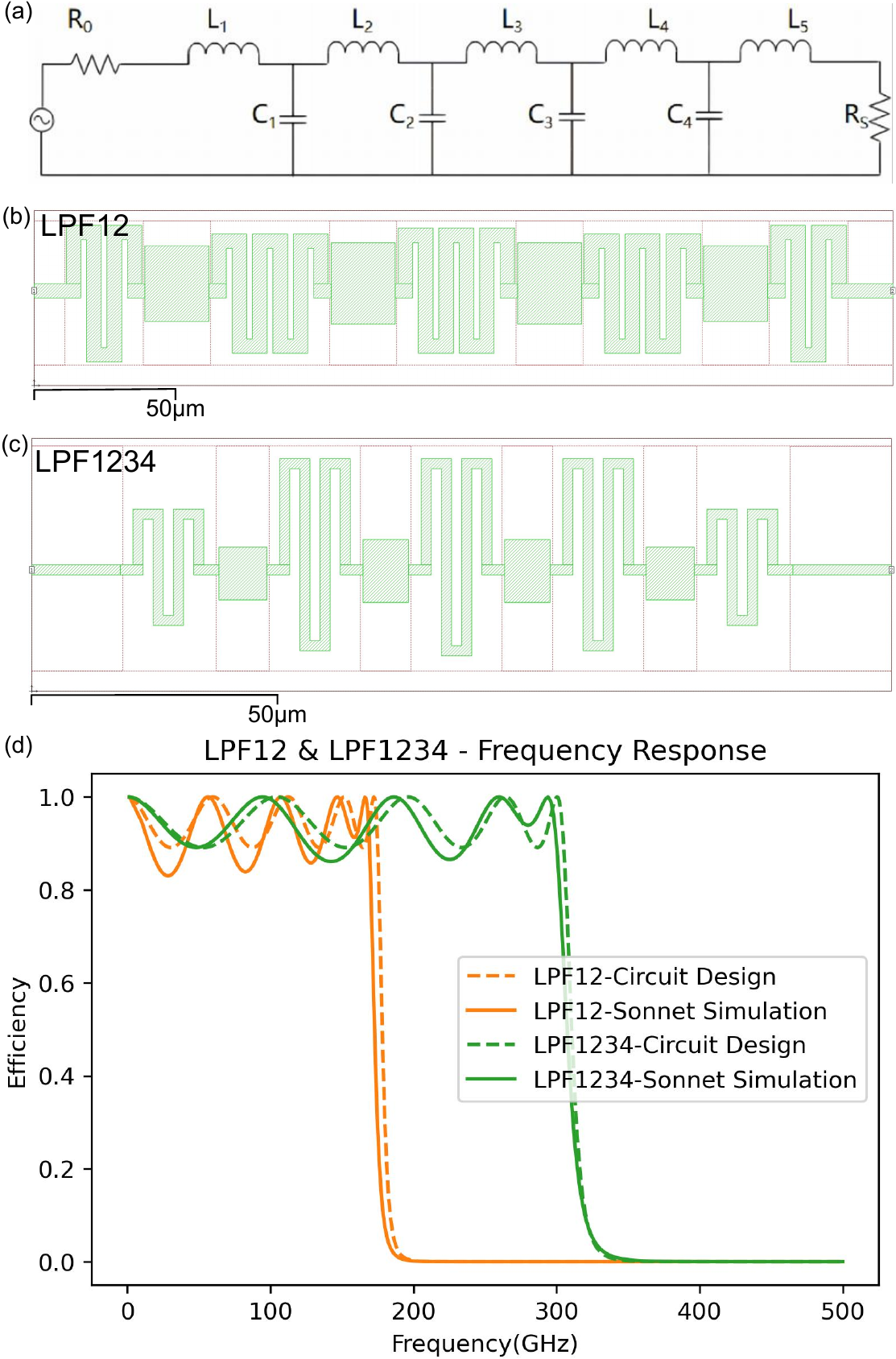}
\caption{The LP filter designs.  (a) Lumped-element schematic.  (b) L12 layout showing meandered inductors and shunt capacitor top plates.  The inductors sit in holes cut out of the ground plane, while the shunt capacitors' other plate is the ground plane. (c) L1234 layout. (d) Power transmittance  ($|S_{21}|^2$) calculated for lumped-element circuits and from Sonnet for the physical layouts, after adjusting the shunt capacitor values as described in the text.}
\label{LPF1234_and_LPF12.pdf}
\end{figure}




\subsection {Band-pass Filter Designs}



\begin{figure}
\centering
\label{tab:bpf5_6}
\includegraphics[width=3.5in]{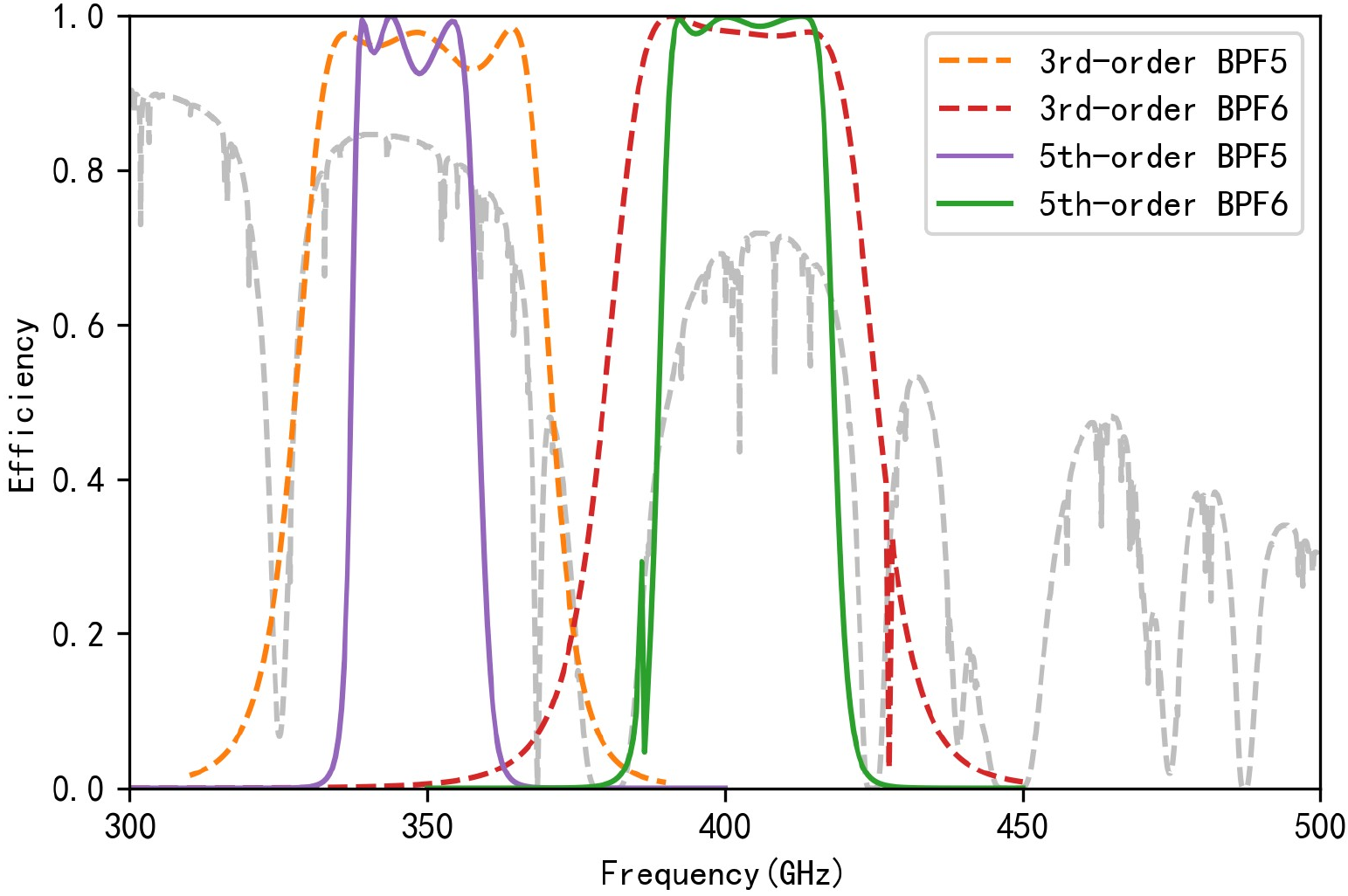}
\caption{Transmittance calculated in Sonnet for optimized 3rd-order and 5th-order bandpass filter designs for B5 and B6.  The narrow atmospheric windows necessitate sharper cutoffs than for B1--B4, which use 3rd-order filters.}
\label{BPF5_and_6.pdf}
\end{figure}

\begin{figure}
\centering
\label{tab:bpf_design_steps}
\includegraphics[width=3.5in]{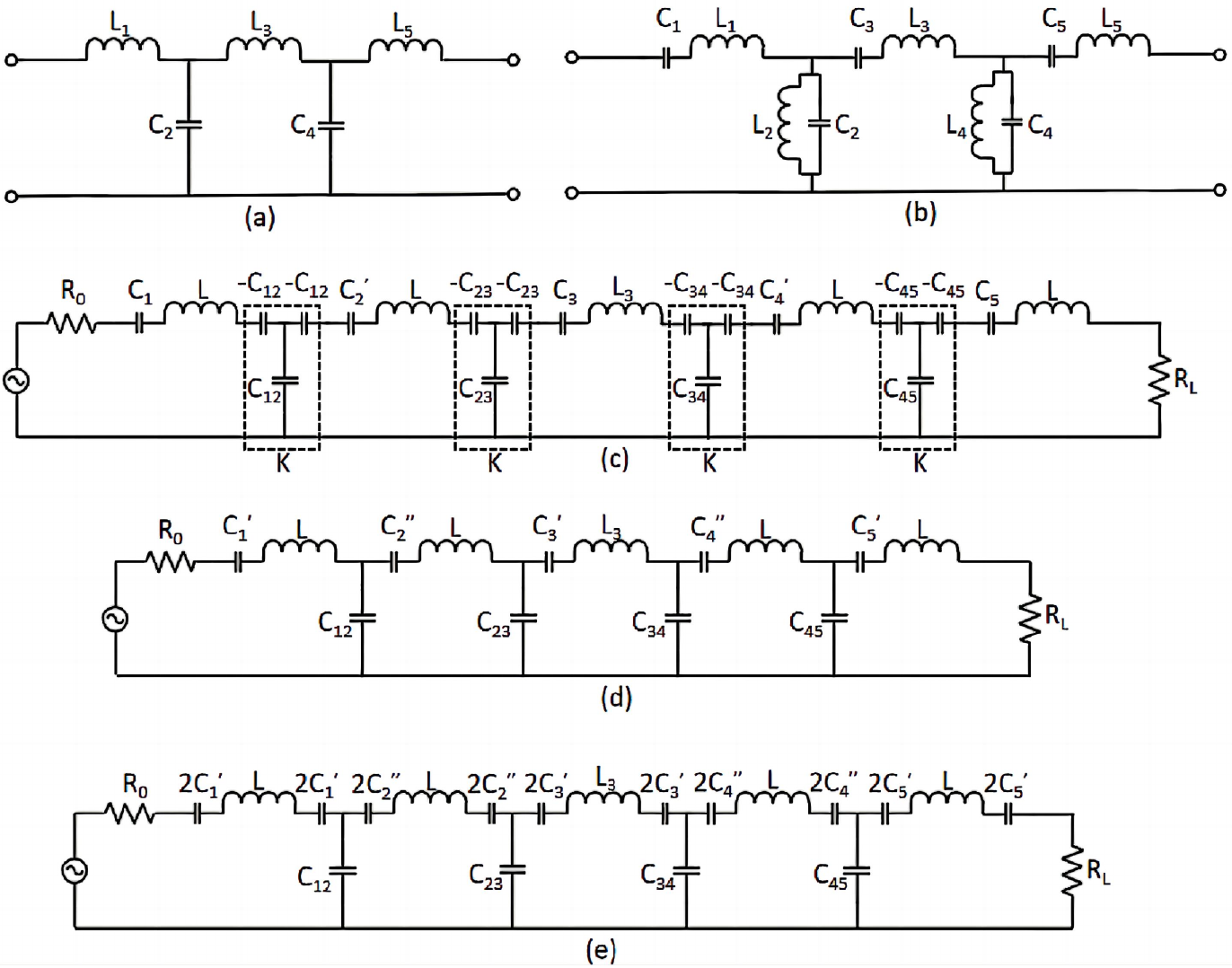}
\caption{Steps for 
5th-order BPF 
design, based on~\cite{IEEEtas2009}. (a)~5th-order (5 element) Chebyshev LPF prototype. 
(b)~Transformation of LPF into BPF by replacing series inductors and shunt capacitances with, respectively, series and parallel LC networks. 
(c) Replacement of shunt inductors with impedance inverters for fabrication simplification.
(d) Absorption of inverter capacitors into series capacitances.
(e) Replacement of single series capacitances by dual series capacitances for fabrication simplification.
}
\label{BPF circuit design steps.pdf}
\end{figure}

We construct band-pass filter designs using the standard procedure for transforming low-pass filters into band-pass filters~\cite{Microwave Engineering} but following the implementation in~\cite{IEEEtas2009}, which obviates difficult-to-fabricate shunt inductors to ground by use of impedance inverters.  For B1--B4, the 3rd-order filters in~\cite{IEEEtas2009} are sufficient.  Figure~\ref{BPF5 & 6.pdf} shows the narrowness of the B5 and B6 atmospheric windows necessitates 5th-order filters.  Figure~\ref{BPF circuit design steps.pdf} shows our procedure, analogous to that in~\cite{IEEEtas2009}, for transforming a 5th-order LPF (5 elements) into a 5th-order BPF (19 elements).  
As in~\cite{IEEEtas2009}, for simplicity we choose the same value for all the inverter capacitors, $C_{12} = C_{45} = C_{23} = C_{34} = \frac{1}{\omega_0 K_{12}}$ where $K_{12} = \sqrt{\frac{L_2}{C_2}}$.  Again following~\cite{IEEEtas2009}, we then absorb the negative inverter capacitances into the series capacitances and, again for simplicity, choose symmetric values for those series capacitances, yielding
\( C_4' = C_2'\), 
\( C_1' = C_5'\), \( C_4'' = C_2''\), and 
\begin{equation}
C_1' = \left(\frac{1}{C_1} - \frac{1}{C_{12}}\right)^{-1},
\end{equation}
\begin{equation}
C_2'' = \left(\frac{1}{C_2'} -\frac{1}{C_{12}}- \frac{1}{C_{23}}\right)^{-1},
\end{equation}
\begin{equation}
C_3' = \left(\frac{1}{C_3} -\frac{1}{C_{23}}- \frac{1}{C_{34}}\right)^{-1}.
\end{equation}
We also replace these series capacitances with dual series capacitances for fabrication simplification (see Figure~\ref{filterbanks.pdf} layouts).
For the inductors, we also take symmetric values, with $L = L_1 = \frac{R_0 g_1}{\omega_0\Delta \omega}$ and $L_3 = \frac{R_0 g_3}{\omega_0\Delta \omega}$ where $g_1$ and $g_3$ are normalized inductances from~\cite{Microwave Engineering} (Table 8.4, 0.5~dB ripple). 

\begin{table}
\centering
\caption{The elements value of the BPF after LPF transformation}
\label{tab:bpf_values}
\setlength{\tabcolsep}{2pt} 
\begin{tabular}{p{1.2cm} p{1.1cm} p{1.4cm} p{1cm} p{1cm} p{1cm} p{1cm}}
\textbf{} & \textbf{BPF1} & \textbf{BPF2} & \textbf{BPF3} & \textbf{BPF4} & \textbf{BPF5} & \textbf{BPF6} \\
\textbf{Element} & \textbf{79-107} & \textbf{133-177} & \textbf{201-246} & \textbf{270-310} & \textbf{335-360} & \textbf{390-411} \\
\textbf{Value} & \textbf{GHz} & \textbf{GHz} & \textbf{GHz} & \textbf{GHz} & \textbf{GHz} & \textbf{GHz} \\
$L$ (nH) & 0.3357 & 0.2089 & 0.1073 & 0.1207 & 0.4018 & 0.4783 \\ 
$C_1'$ (pF) & 0.0116 & 0.0067 & 0.0056 & 0.0028 & 0.00055 & 0.00034 \\ 
$C_{12}$ (pF) & 0.0388 & 0.0233 & 0.0312 & 0.0240 & 0.01052 & 0.00912 \\ 
$C_2''$ (pF) & 0.0165 & 0.0094 & 0.0069 & 0.0032 & 0.00058 & 0.00036 \\ 
$C_3'$ (pF) &        &        &        &        & 0.00037 & 0.00023 \\ 
$L_3$ (nH) &        &        &        &        & 0.5985 & 0.7125 \\
\end{tabular}
\end{table}

Among the six BPFs,  BPF3 and BPF4 have an input impedance of \( 19 \, \Omega \), while the others have an input impedance of \( 37 \, \Omega \). Following determination of cutoff frequencies \( \omega_1 \) and \( \omega_2 \), the component values for these filters can be calculated prior to the design phase.


As for the LPF, we simulate the individual capacitors and inductors in Sonnet first and iterate to obtain the desired component values, and then we combine them
to form a BPF. As for the LPF, after the combination, the sizes of the parallel plate capacitors and the lengths of the inductors need to be optimized to achieve the desired passband response. Table 1 shows the component values before the 
BPF optimization step. 

We found that three specific component changes were most useful for the above optimization. First, increasing the shunt capacitance to ground enhances the in-band transmittance but reduces the bandwidth. Second, decreasing the inductance increases the cut-off frequency, with a comparatively smaller negative impact on transmittance than doing so by reducing capacitance. Third, adjusting the inductance and capacitance together provides flexible bandwidth control: simultaneously decreasing inductance and increasing capacitance effectively widens the bandwidth, whereas increasing inductance and decreasing capacitance narrows it.

\subsection {Filterbank}
After optimizing each individual filter, they are combined into filterbanks. However, the performance of the filterbanks does not immediately meet the target specifications because the filters affect each other via parasitic mutual couplings. 
Even though we selected T-section filters to minimize the \textit{direct} impact of the filters on one another, there are mutual couplings between the inductors of the different filters because they are close to each other.  The coupling appears to be via the ground plane: if we minimize the ground plane between adjacent inductors, we minimize this parasitic coupling.  Therefore, the layouts maximize the empty space between adjacent inductors of different filters.


\begin{figure*}
\centering
\includegraphics[width=\linewidth]{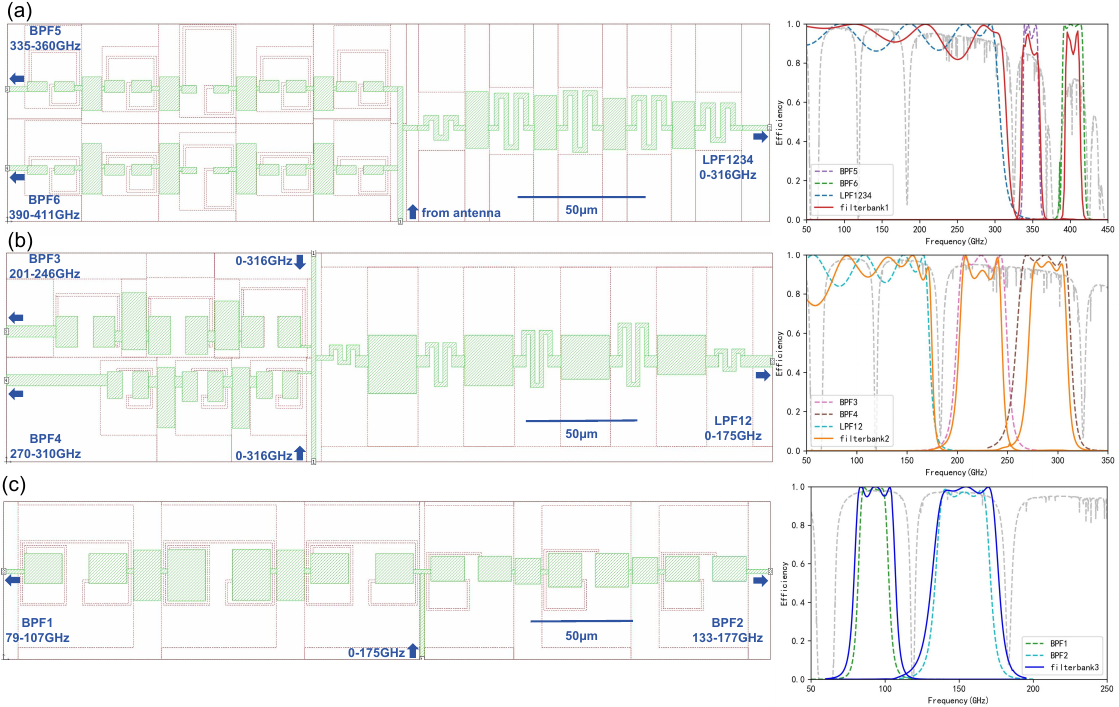}
\caption{Left: filterbank layout.  Right: transmission calculated in Sonnet.  Dashed lines show calculations for each filter individual while the solid lines show the transmission of the assembled filterbank.  Top: LPF1234-BPF5-BPF6.  Middle: LPF12-BPF3-BPF4.  Bottom: BPF1-BPF2.}
\label{filterbanks.pdf}
\end{figure*}

The LPF1234-BPF5-BPF6 is the first filterbank connected to the output port of the level 0 antenna array transmission line. Port 1 (\( 37 \, \Omega \)) connects to the antenna's output port, while LPF1234, BPF5, and BPF6 output to ports 2, 3, and 4 respectively, all \( 37 \, \Omega \).
Because the LPF12-BPF3-BPF4 filterbank is also a summing junction, its input ports (1U and 1L) are 37 Ohms but its output ports 2, 3, and 4 (LP12, B3, and B4) are all \( 19 \, \Omega \).  All three output microstriplines are tapered back down to \( 37 \, \Omega \) impedance before going to detectors or the next filterbank.
Finally, the low-pass-filter LPF12 output of the above filterbank is received at port 1 (\( 37 \, \Omega \)) of the BPF1-BPF2 filterbank, which outputs to port 2 and port 3 for B1 and B2 (both \( 37 \, \Omega \)).


\begin{figure}
\centering
\includegraphics[width=3.5in]{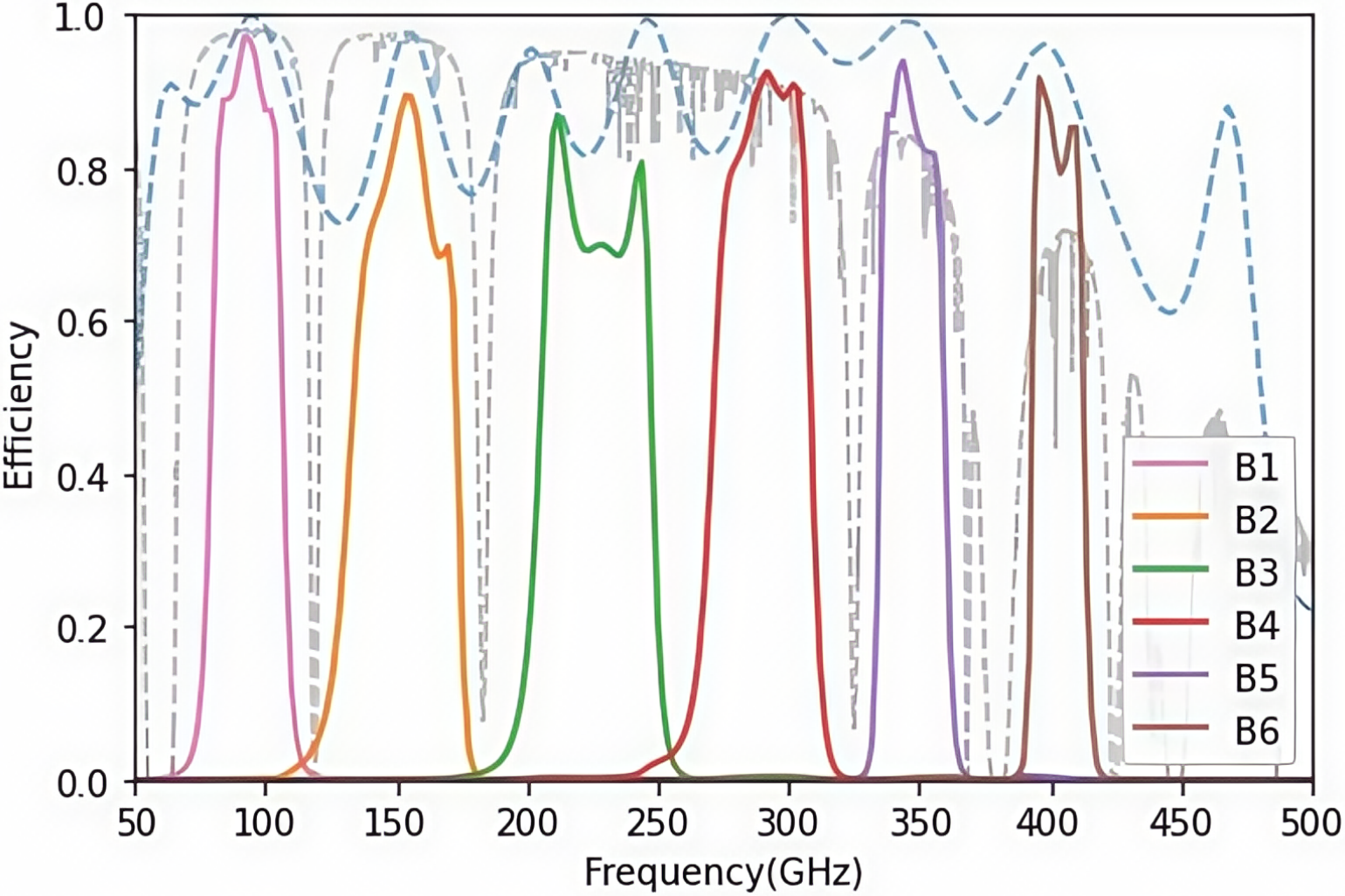}
\caption{The 
product of the transmittances of the antenna and filterbanks 
after optimization.  Also shown are the atmospheric transmittance (1~mm PWV) and antenna transmittance alone.}
\label{efficiency_of_B1_to_B6.pdf}
\end{figure}

Fig. \ref{filterbanks.pdf} shows the optimized filterbank layouts and the transmittances before and after combining and reoptimizing the filters. 


\section{Design Results and Discussion}

Fig. \ref{efficiency of the B1-B6.pdf} shows the full antenna and filter network performance obtained by multiplying the transmittance of the antenna and the filterbanks relevant for each band.  In spite of the large number of cascaded elements, the overall expected transmittance is quite good, with the mean transmittance across all bands better than 80\%.

We have achieved this excellent performance by thoughtful choice of filter type and careful optimization.  We began the design with T-type low-pass filters with inductive input components to minimize direct capacitive coupling between filters.  We minimized the inductor lengths to decrease parasitic capacitance.  We obtained further improvements to individual filter transmittances by carefully making coordinated changes to the inductance and capacitance values. When combining filters into filterbanks, we minimized the length of the connecting microstriplines to minimize impedance transformation effects and adjusted the ground plane fill geometry to mitigate the effect of parasitic mutual couplings.



\begin{figure}
\centering
\includegraphics[width=3.5in]{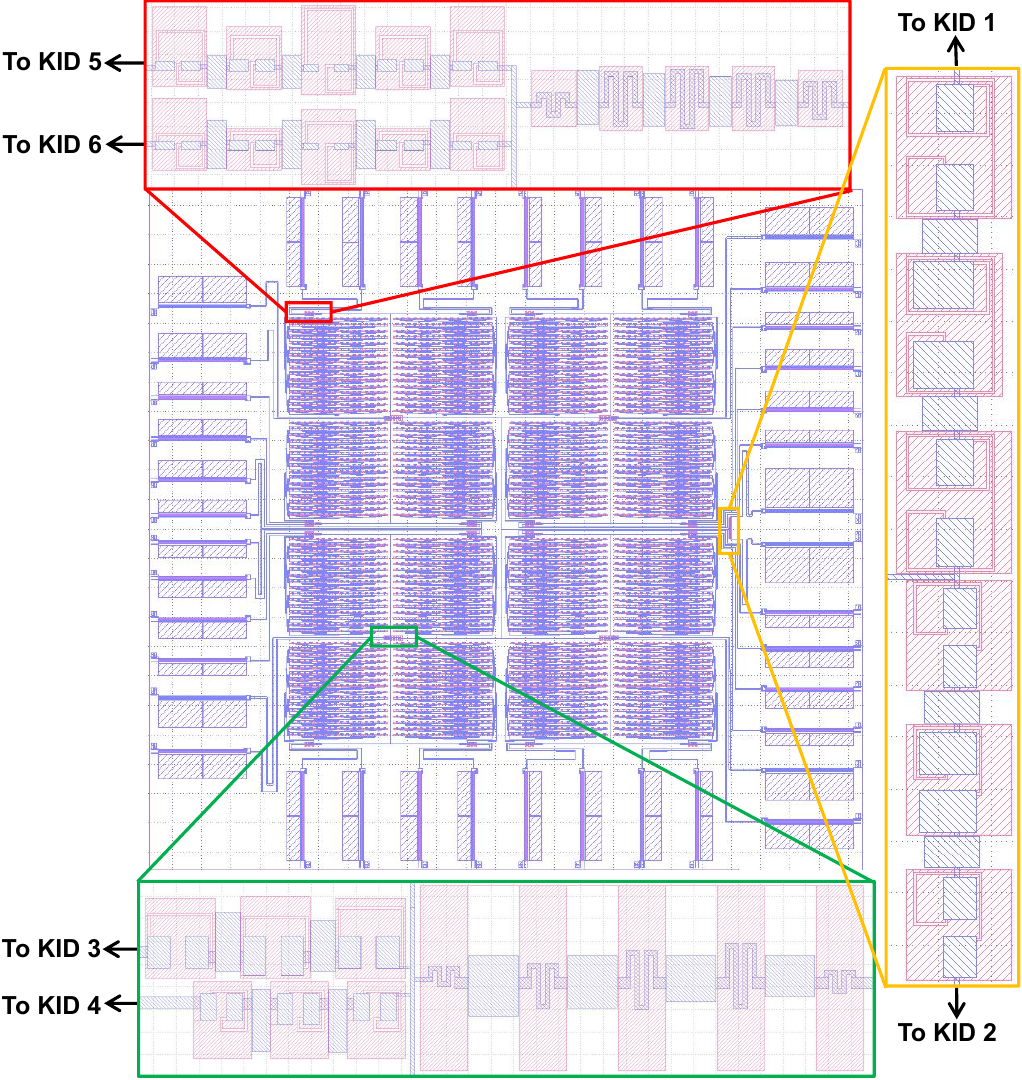}
\caption{The layout of 75-420GHz kinetic inductance detector pixels for NEW-MUSIC. (a) Layer structure. \cite{jltp2022}. (b) These structures couple trans-mm power from the microstripline entering at the left into the KID inductor. The Nb microstrip line top layer is shown in blue. The bottom blank area is the grounded niobium, and the pattern etched away is shown in red.}
\label{pixe_layout.pdf}
\end{figure}

Our design achieves the goal of a six-band, 80-420 GHz (2.4-octave) pixel for NEW-MUSIC.  We show in Fig. \ref{pixel layout.pdf} the full layout of the antenna with the filterbanks integrated.  The various band-pass filter outputs terminate in microstripline-coupled, parallel-plate-capacitor, lumped-element KIDs (MS-PPC-LEKIDs) whose design was described in \cite{mkid2004}.  The antenna/filterbank design is, however, generic: any microstripline-coupled detector could be used.


\section{Conclusion and future plans}


To address the need for multi-band astronomical observations, we have implemented a hierachically summed phased array antenna of slot dipole antennas with integrated filterbanks to enable the frequency-selective summing. The fundamental unit is a 16-element array of slot dipole antennas, each with 16 feeds, with all 256 feeds combined via an equal-path-length binary summing tree, which provides good transmittance over a nearly 10:1 (3.3 octave) bandwidth, 50--500 GHz. Three filterbanks, using six band-pass filters and two low-pass filters, enable frequency-selective hierarchical summing and band definition consistent with atmospheric transmission windows. We use Sonnet to optimize the component choice for the individual filters and then to re-optimize the component values after combining the filters in to filterbanks.

Our next step is to integrate the antenna/KID design shown in Fig. \ref{pixel layout.pdf} into a test device design with four such pixels as well as a number of diagnostic structures that measure microstripline loss, wavespeed, and impedance.  We will measure optical efficiency, spectral bandpasses, and radiation patterns to test the performance of the antenna and seek feedback for refinement of the design.  We plan to deploy a 4$\times$4 array of such antennas in the NEW-MUSIC instrument in 2027.

\section*{ACKNOWLEDGMENT}
Xiaolan Huang thank Dr. Xu Yu for useful discussion during simulation of the filter.



\begin{IEEEbiography}[{\includegraphics[width=1in,height=1.25in,clip,keepaspectratio]{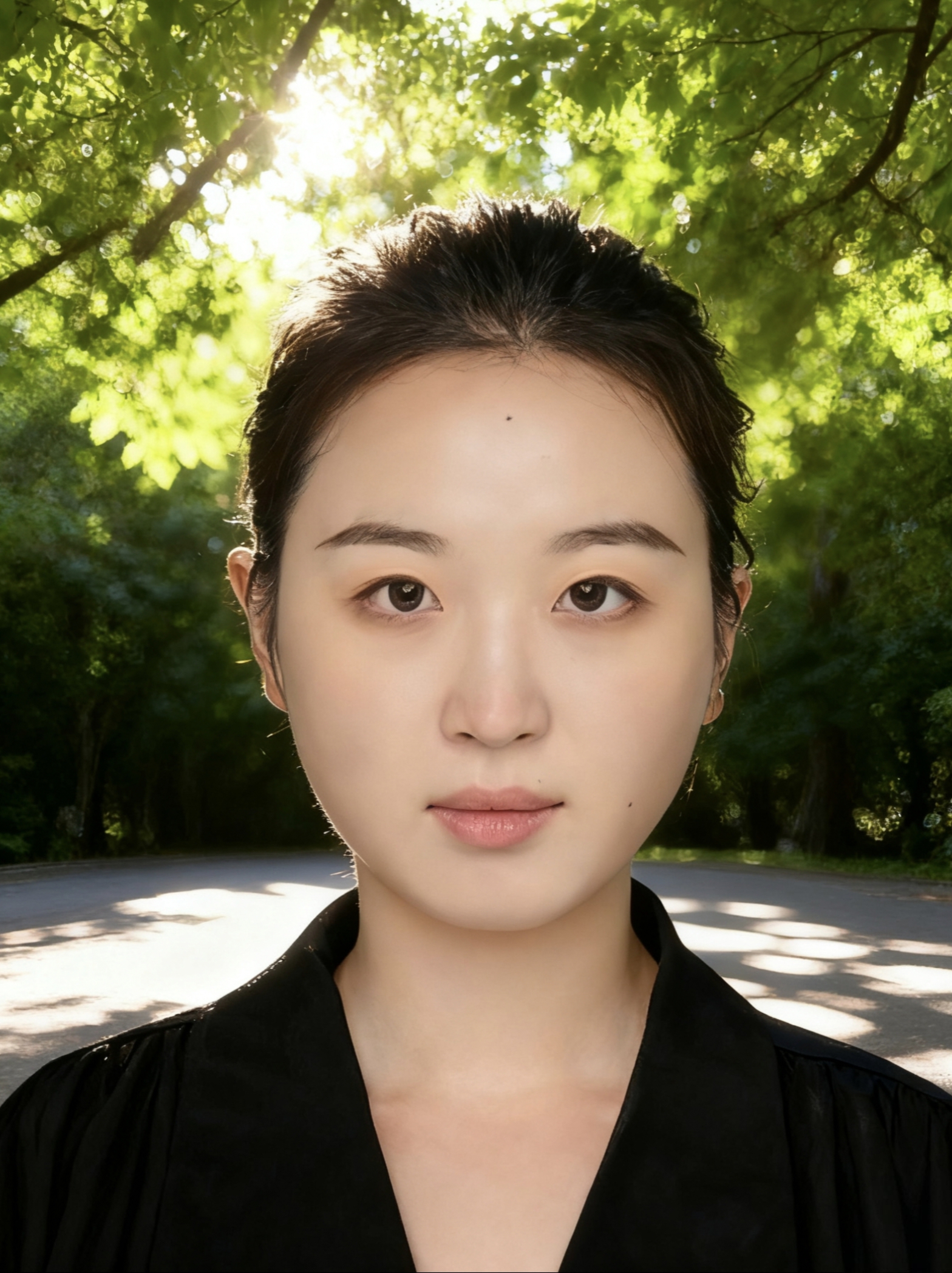}}]{Xiaolan Huang} is currently a Ph.D. candidate in Astronomical Detection Technology at Shanghai Normal University, a program she began in 2022. Her primary research focuses on the design and development of superconducting detectors, specifically aimed at enhancing their sensitivity and efficiency for applications in millimeter-wave and submillimeter-wave astronomy. Her work involves the design and fabrication of kinetic inductance detectors (KIDs) and the optimization of cryogenic readout systems. Her research interests will also extend to the integration of these advanced detectors into new telescope instrumentation for cosmic microwave background (CMB) observations and the study of star-forming galaxies.\end{IEEEbiography}

\begin{IEEEbiography}[{\includegraphics[width=1in,height=1.25in,clip,keepaspectratio]{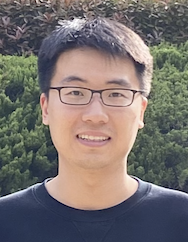}}]{Shibo Shu} received the Ph.D. degree in physics from the University of Grenoble Alpes / Institut de Radioastronomie Millimétrique (IRAM) in 2019. He is currently an associate researcher at the Institute of High Energy Physics, CAS and an associate professor at the University of Chinese Academy of Sciences. His research focuses on cutting-edge high-sensitivity detection technologies and ultra-low-noise amplification, for astronomical observations. He is currently engaged in the development of transition edge sensors and kinetic inductance detectors for multiple projects.\end{IEEEbiography}

\begin{IEEEbiography}[{\includegraphics[width=1in,height=1.25in,clip,keepaspectratio]{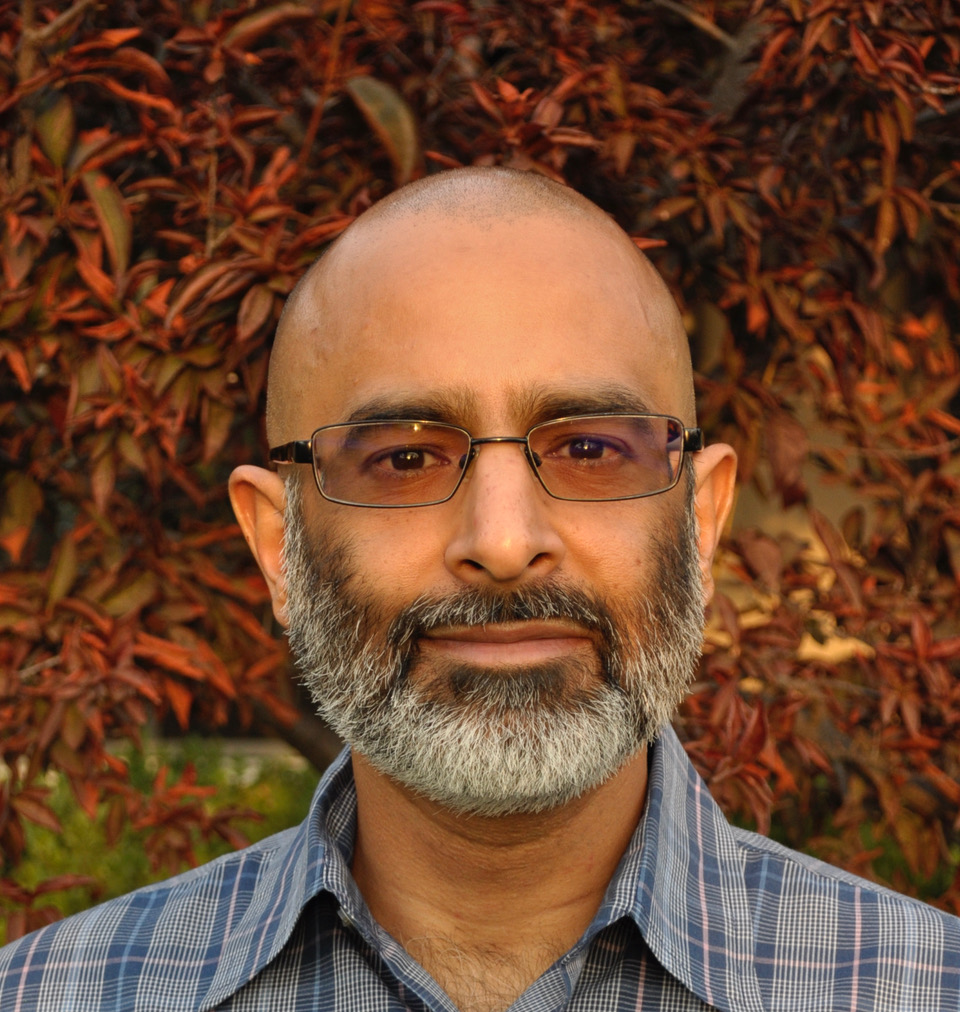}}]{Sunil R. Golwala} received the Ph.D. degree in physics from the University of California, Berkeley, CA, USA, in 2000.
He is currently a Professor of Physics with the California Institute of Technology (Caltech), Pasadena, CA, USA. His interests are in the development of new instrumentation and telescopes for astronomical measurements at millimeter and submillimeter wavelengths, particularly for the cosmic microwave background (CMB), the study of hot plasmas using
the Sunyaev–Zel’dovich effect, dusty star-forming galaxies, and for time-domain sources. His work includes the development of silicon metamaterial optics, superconducting phased-array antennas and bandpass filters, and microstripline coupled kinetic inductance detectors (KIDs).\end{IEEEbiography}

\begin{IEEEbiography}[{\includegraphics[width=1in,height=1.25in,clip,keepaspectratio]{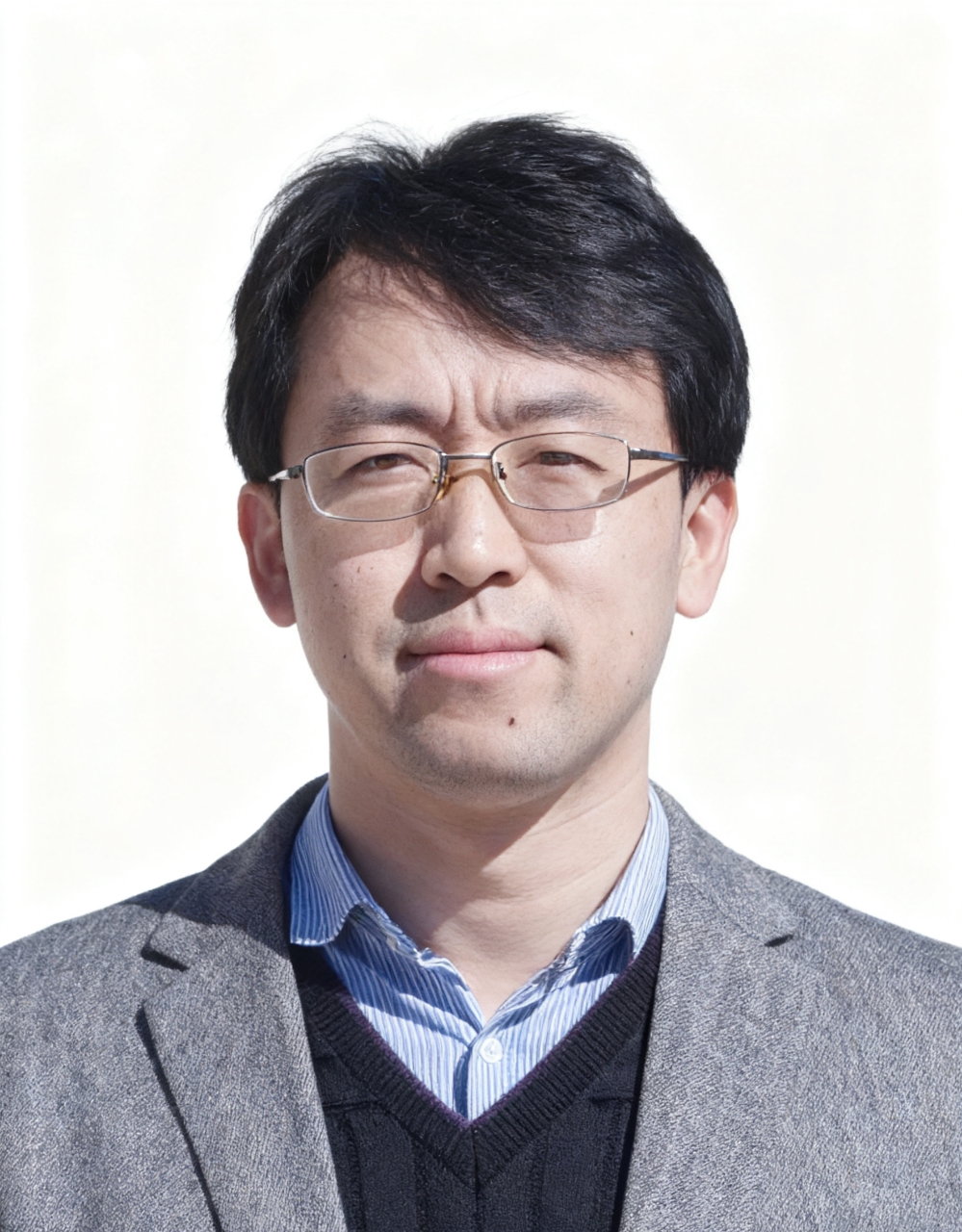}}]{Fengliu} received his Ph.D. in Physics from Fudan University, Shanghai, China, following B.S. and M.S. studies in the same discipline at the university. He conducted postdoctoral research in the Department of Materials Science and Engineering at the University of Pennsylvania, Philadelphia, PA, USA.

His research spans three main areas:

1. Terahertz Astronomical Detection Technology and Device Development—playing a key role in establishing the Shanghai International Joint Laboratory for Submillimeter Astronomy. His work focuses on the development of Nb/NbN superconducting films, tunnel junctions, and superconducting mixers/receivers for frequencies such as 230/460 GHz and 345/650 GHz, along with maintenance and upgrades for the LCT submillimeter telescope receiver.
2. Plasmonics, Photonic Crystals, and Bio-inspired Optics
3. Optical and Optoelectronic Properties of Low-Dimensional Materials such as graphene, and related micro/nano-optoelectronic devices.

He has authored over 100 SCI-indexed publications, which have received more than 2,000 citations, and holds 11 national invention patents. He has led numerous research projects supported by the National Natural Science Foundation of China (including Key, General, and Youth Programs), as well as major science and technology initiatives from Shanghai Municipality.\end{IEEEbiography}

\end{document}